\documentclass[fleqn,twoside]{article}
\usepackage{espcrc2}

\usepackage{epsfig}
\usepackage{espcrc2}

\title{Coherent Neutrino--Nucleus Scattering}
\date{}

\author{E.A. Paschos\address{Universit\"at Dortmund,\\
        D--44221 Dortmund, Germany}
        and
        A. Kartavtsev\address{Rostov State University,\\
        Rostov--on--Don, Russia}\\
       (presented by E.A.\ Paschos)}

\begin{document}

\begin{abstract}
We review coherent scattering of neutrinos on nuclei by coupling
the weak currents to vector and axial-vector meson states. The
couplings are obtained from known reactions, like $\tau$--lepton
and vector meson decays.  We compute and present the cross
sections as functions of energy and momentum transfer for various
kinematic regions, including those which are relevant to
oscillation experiments.  The results are presented in figures and
are consistent with numerical values used in the interpretation of
oscillation experiments. \vspace{1pc}
\end{abstract}

\maketitle

\section{Introduction}
Coherent scattering occurs when photons or neutrinos interact with
two or more particles and the amplitudes from the various
particles in the target add up.  Coherent phenomena have been
observed in electromagnetic interactions when photons interact
with nuclei or atoms or layers of atoms, as is the case with Bragg
scattering.  A consequence of coherence is an increase of the
cross--section becoming proportional to the square of the number
of particles in the target leading to increased counting rates.
A reaction where coherent scattering has been reported
\cite{ref4}--\cite{ref11} is

\begin{equation}
\nu + {\rm nucleus} \to \mu^- +\pi^i +X^i\, .
\label{reaction}
\end{equation}

A basic condition for coherent scattering requires the wavelength
of the incident particle or the wavelength of the momentum
transferred to the target to be large enough so that the entire
region within the wavelength contributes to the amplitude. This
situation is difficult to realize in neutrino reactions because
the radius of the nucleus sets the scale
\begin{equation}
R_{\rm nucleus} \approx 2\,{\rm fm} \approx \frac{1}{100\,{\rm
MeV}}
\end{equation}
and it is fulfilled either at very low neutrino energy where the
cross sections are small or low momentum transfers $|t|$
\raisebox{-0.1cm}{$\stackrel{<}{\sim}$} 0.10 GeV$^2$ where the
phase space is limited.  In the latter case it is also necessary
to define the particles and the method for controlling the
momentum transfer, so that we know that the events originate from
coherent scattering.

Several cases of coherent neutrino scattering have been discussed
in articles \cite{ref1}-\cite{ref2} but only the reaction in eq. 
(\ref{reaction}) has been observed experimentally. The main
reason is that the cross--sections are very small. 

A first case of coherent scattering occurs when the momentum transfer between 
the incident neutrino and the scattered lepton is small. This corresponds 
to the parallel configuration of the leptons. At $Q^2=0$, the PCAC 
hypothesis allows one to derive the contribution of the axial current to the cross
section \cite{ref3}, \cite{ref14}:

\begin{equation}
\frac{d\sigma}{d \nu d Q^2 d k^2_T}\biggl|_{Q^2=0} = 
\frac{G_F^2}{2\pi^2}
\frac{E'}{E}\frac{f_{\pi}^2}{\nu}\frac{d \sigma^{\pi N}_{el}}{d k^2_T}
\end{equation}

\noindent The extension to non-zero $Q^2$ is given by other
mesons, like $A_1$ or $\rho\pi$ pairs \cite{ref14}.

A second case of coherent scattering occurs when the wavelength of the momentum
transferred to the target nucleus is large enough. This is satisfied when the 
4--momentum transfer squared to the nucleus is small. We shall denote 
the momentum transfer by t and for $|t|<0.10$ GeV$^2$  a peak has been observed 
for reaction (\ref{reaction}) in several experiments, which is interpreted as the coherent
scattering from the nucleus.  In the zero--recoil approximation
\begin{equation}
\begin{array}{l}
t= (q-p_{\pi})^2 =
\\[2mm]
\hspace*{10mm}-\left[ \sum_{\mu,\pi} (E_i-p_i^{\|}) \right]^2-
\left[ \sum_{\mu,\pi}p_i^{\perp}\right]^2
\end{array}
\end{equation}
where the right--hand side depends on the transverse and parallel
momentum components of the muon and the pion, both of which are
detectable in the experiments \cite{ref13}.

An additional requirement for coherent scattering demands the wave function of
the nucleus to be the same before and after the interaction.
This puts restrictions on the intermediate mesons interacting with
the nucleons. The scattering on the nucleus should not change
charge, spin, isospin or other quantum numbers. Since the  isospin
operator, $\tau_3$, has eigenvalues with opposite sign for proton and neutron, the
contribution of the isovector intermediate mesons is proportional
to difference of number of protons and neutrons and therefore
negligible for most nuclei. Consequently, in order for proton and
neutron contributions to add up the intermediate mesons must be
isoscalars.

Diagrams which are allowed are shown in figure 1.
\begin{figure}[h]
\begin{tabular}{l}
\epsfig{file=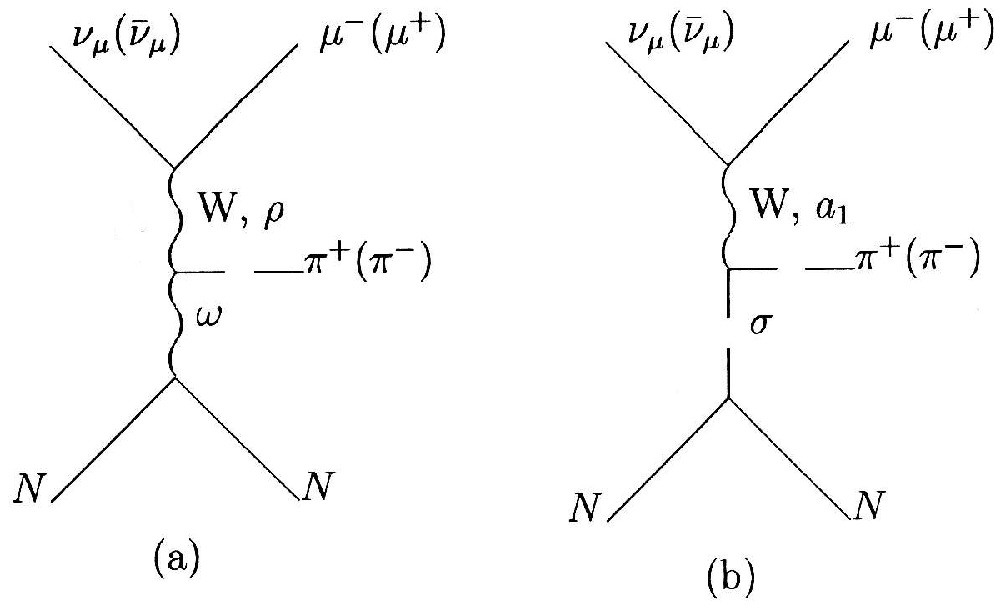,width=7cm}\\
\epsfig{file=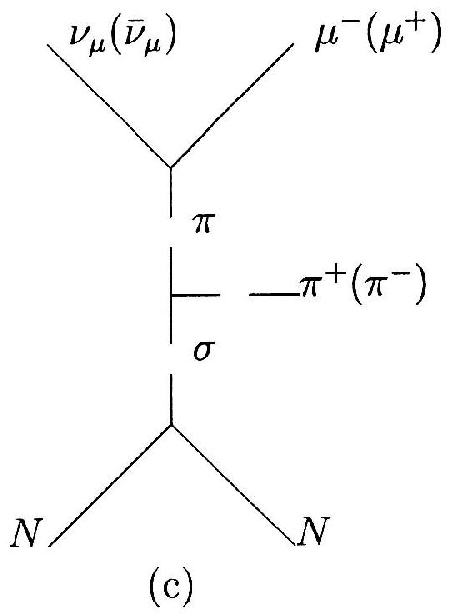,width=3.5cm}
\end{tabular}
\caption{Leading channels contributing to coherent single pion
neutrino production.}
\end{figure}
\setcounter{figure}{1}

\noindent According to these diagrams intermediate vector boson
is replaced  with the $\rho$ and the $A_1$ mesons and the pion.
The remaining part of the 
diagrams is computed as the scattering of hadronic
states.  Piketty and Stodolsky \hspace*{1pt} \cite{ref12}
estimated the production of $\pi,\, \rho,\, A_1$--mesons
\mbox{using} meson--dominance, suggesting that the vector current
is replaced by the $\rho$--meson and the axial current by $A_1$
and other mesons.  Rein and Sehgal \cite{ref13} developed a model
with intermediate mesons, introducing a form factor for the
nucleus in order to describe coherence. The last active group
in this field is Belkov and Kopeliovich \cite{ref14} who suggest
that a summation over intermediate meson states is more accurate.
These suggestions are appealing but it is still hard to compute
them accurately\cite{Kopeliovich}.  We shall describe a calculation along these
lines, which relies on the available data and estimates the
coherent cross--section at lower and higher energies.

\section{Effective Couplings}
The effective Lagrangian for the low--energy weak interaction of a
charged vector meson is \cite{Lichard}
\begin{equation}
\mathcal{L}_\rho = -\frac{g_W}{2}\, \frac{f_\rho}{2}\,
 V_{ud}\, W_{\mu\nu}\, \rho^{\mu\nu}
\end{equation}
with $W_{\mu\nu}=\partial_{\mu}W_{\nu}-\partial_\nu W_{\mu}$ and
$\rho^{\mu\nu} =\partial^{\mu}\rho^{\nu}-\partial^{\nu}
\rho^{\mu}$. Whenever the $\rho$ is on the mass--shell the
Lagrangian becomes
\begin{equation}
\mathcal{L}_\rho= -\frac{g_W}{2}V_{ud}\, f_\rho\,
  m_{\rho}^2\, W_{\mu}\rho^{\mu}\, .
\end{equation}
In these equations $g_W$ is the SU(2) coupling constant and
$V_{ud}=\cos(\theta_c)$ with $\theta_c$ the Cabibbo angle.

Similarly the effective interaction of a charged axial meson on
the mass shell is given by
\begin{equation}
\mathcal{L}_{A} = \frac{g_W}{2} V_{ud} f_A
  m_A^2 W_{\mu} A_1^{\mu}\, .
\end{equation}
The constants $f_{\rho}$ and $f_{A_1}$ are coupling constants of
the respective vector mesons to the $W$--boson and must be
determined by experiments. The decays of the $\tau$--lepton
provide this opportunity. Combining with the Standard Model Lagrangian we
easily obtain Lagrangian of effective interaction of the charged
current with the $\rho$ meson:
\begin{equation}
\displaystyle {\cal L}_{eff}=G_F f_\rho m_\rho^2 V_{ud}
\bar{e}\gamma^\mu (1-\gamma^5)\nu \rho_{\mu}
\end{equation}
and there is a similar formula for the $A_1$--meson.

The width of the process $\tau\rightarrow\nu_\tau+\rho$  is \cite{Lichard}
\begin{equation}
\begin{array}{l}
\displaystyle \Gamma(\tau\to \nu+\rho) = (G_FV_{ud}f_\rho)^2
\frac{m_\rho^2}{8\pi m_{\tau}^3}\nonumber
\\[2mm]\displaystyle\hspace*{20mm}
\left(m_{\tau}^2-m_\rho^2\right)^2
\left(m_{\tau}^2+2m_\rho^2\right).
\end{array}
\end{equation}
This gives $f_\rho\approx 0.16$. Since chiral symmetry is broken,
relation $f_a=f_\rho$ is modified and analysis of experimental
data \cite{Lichard} gives $f_a\approx0.8\cdot f_\rho$.

The next coupling refers to the decay $ A_1 \to \pi +\sigma $ with
the effective coupling
\small
\begin{equation}
g_1^{a_1\sigma\pi} \sigma a_{1\mu}\partial^{\mu}\pi+g_2^{a_1\sigma\pi}
a_1^{\mu\nu} (\partial_{\mu}\pi\partial_{\nu}\sigma -
 \partial_{\nu}\pi\partial_{\mu}\sigma)\, .
\end{equation}
\normalsize
The total width of the $A_1$ is poorly known (being very wide) and
the branching ratios are not determined. For that reason values of
$g_1^{a_1\sigma\pi}$ and $g_2^{a_1\sigma\pi}$ were determined in \cite{aisipi} from theoretical
considerations based on extended Nambu--Jona-Lasinio model.

The amplitudes for the various diagrams are written following
standard Feynman rules. They also involve the couplings of
isoscalar mesons to the nucleus for which the contributions of
protons and neutrons add up. When the momentum transfer to the
nucleus is large, the scattering is incoherent.  For small
momentum transfers $|t|<0.10$ GeV$^2$ the whole nucleus participates
and contributes coherently.  This aspect is included
phenomenologically by introducing a factor $F(t)=e^{-b|t|}$ with
$b$ being determined experimentally.

\begin{figure}[Htb]
\epsfig{file=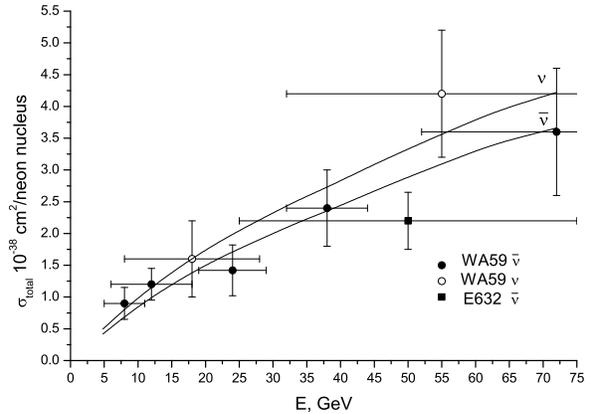,width=8.5cm} \caption{Total cross
section of coherent single pion neutrino and antineutrino
production.}
\end{figure}

The experiments distinguish two categories of events: events with
and without stubs.  Events with stubs are incoherent since the
nucleus breaks up -- they have no special characteristics.  For
events without stubs, the experiments observed a large peak at
$|t|<0.10$ GeV$^2$. Many experiments attempt to determine $b$ but
they reported a large range of values.  The experimental groups
also report \cite{ref10,ref11} the energy dependence of the
coherent pion production shown in fig.\ 2.

 We use the sum of the diagrams described above to
produce the curves presented in fig.\ 2 and obtained the value
$b=52$ GeV$^{-2}$ for a neon nucleus. We also computed the
differential cross section shown in fig.\ 3.

At $t=0$ the $\rho$--exchange graph vanishes whereas $A1$--exchange 
graph is finite. With increase of  $|t|$ the differential 
cross section ${\rm d} \sigma / {\rm d} |t|$ corresponding to the 
$\rho$-exchange graph grows faster than that corresponding to the
$A1$--exchange. Thus diagram in fig. 1(a) substantially contributes 
to the total cross section. We wish to note here, that in our calculations
we take both longitudinal and transversal polarizations of the gauge
bosons into account.

\begin{figure}[Htb]
\epsfig{file=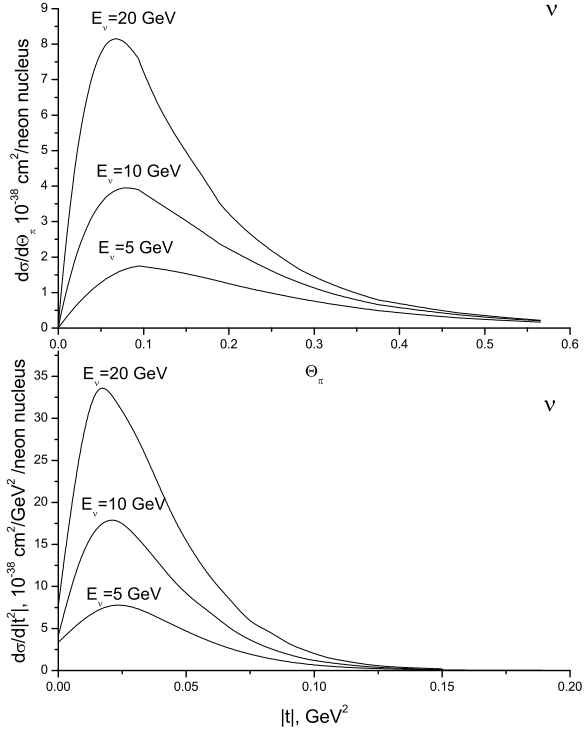,width=8.5cm}
\caption{Differential cross sections ${\rm d}\sigma / {\rm
d}\Theta_\pi$ and ${\rm d} \sigma / {\rm d} |t|$.}
\end{figure}

The cross section peaks at low values of $t$ and
$\theta$ as expected.  The differential cross section falls off
with increasing $|t|$ which is a consequence of the nucleus' form
factor.  The value of $b$ is much larger than typical values of
diffractive scattering, where $b\sim 3-5$ GeV$^{-2}$, which is a
strong indication that it reflects the particle density within the
nucleus.

\section{Numerical Results}
The scattering of neutrinos through charged current receives
contributions mainly from the $A_1$ and $\rho$ intermediate states. 
Contribution of diagram in Fig. 1(c) is proportional to the lepton mass
and therefore negligible.

The
difference between neutrino and antineutrino cross sections is also
determined by the interference between the diagrams 1(a) and 1(b)
and is relatively small.

The integrated cross sections for $CC$ and $NC$ coherent
scattering are depicted in fig.\ 4.

The cross-section per Oxygen nucleus at $E_{\nu}= 2$ GeV
is $0.16\times 10^{-38}$ cm$^2$ and for antineutrinos $0.13 \times
10^{-38}$ cm$^2$. The cross--section for resonance production at
the same energy per Oxygen nucleus is $\sim 5\times 10^{-38}$
cm$^2$.  Thus coherent scattering is $\sim 3\%$ of the charged
current resonance production.

Similar calculations were performed for the scattering of neutral
currents on an Oxygen target. The cross--sections are smaller and
are shown in fig.\ 4.  Again in the case of neutral currents the
coherent production of pions is $3\%$ of resonance production.
The two numerical estimates for coherent scattering add a small
contribution to the neutrino oscillation data at small values of
$|t|$ and agree with the values introduced in the analysis of the
data \cite{Sakuda}.

\begin{figure}[h]
\epsfig{file=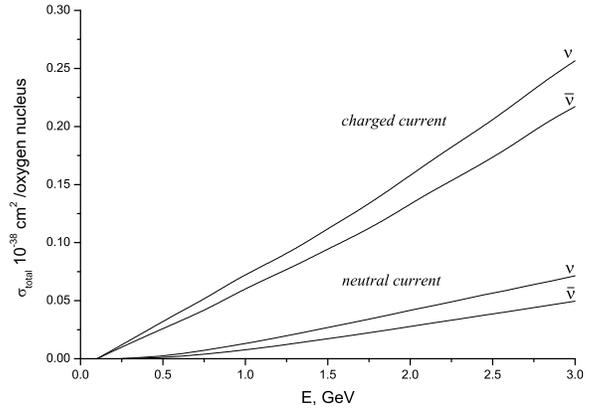,width=8.5cm} \caption{Total cross
section for CC and NC coherent single pion antineutrino and
neutrino production.}
\end{figure}

\section{Summary and Outlook}
Coherent production of pions by neutrinos has been observed in
several experiments.  The events have a characteristic
$t$--dependence which suggests the scattering from the entire
nucleus.  In addition to the $t$--dependence coherent scattering
must show characteristic dependence
\begin{itemize}
\item[(i)] on the momentum transfer between the leptons, $Q^2$,
\item[(ii)] an azimuthal dependence \\
$d\sigma= C_1 + C_2\cos\phi + C_3\cos 2\phi$\\
with $C_1,\, C_2,\, C_3$ functions of momenta etc. independent of
$\phi$, and \item[(iii)] a dependence on the atomic number of the
nucleus $A$, which has not been determined yet.
\end{itemize}
The cross sections at various neutrino energies are correlated
(see figs.\ 2 and 4) and it will be a challenge to observe and
determine properties of the effect at lower energies in the new
generation of neutrino experiments. 

A closely related process is  \cite{Sehgal} -- \cite{Bernabeu}
\begin{equation}
\nu+e^-\to e^-+\gamma + X_1 \;.
\end{equation}
This is a weak process with the pion being replaced by the
emission of a photon and it can be calculated accurately.  The
cross--section is expected to be very small  \cite{Sehgal} --
 \cite{Bernabeu}.

Finally, another interesting process is
\begin{equation}
a+N\to \gamma + X_2
\end{equation}
where $a$ is the field for an axion. Coherent scattering in this
reaction is very important because it  significantly enhances the chances for observing
axions coming from Sun or another extraterrestrial source.

\section{Acknowledgement}
The support of BMBF under contract 05HT1PEA9 and of DFG is
gratefully acknowledged. We thank Dr. Sehgal for helpful correspondence.

\end{document}